\documentclass[prb,twocolumn,preprintnumbers,amsmath,amssymb,showpacs]{revtex4}

\usepackage{graphicx}
\usepackage{dcolumn}
\usepackage{bm}
\usepackage{mathrsfs}
\usepackage{subfigure}
\usepackage{natbib}
\usepackage{color}
\usepackage[normalem]{ulem}

\begin{document}
\title{Soft vortex matter in a type-I/type-II superconducting bilayer}

\author{L. Komendov\'{a}}
\email{lucia.komendova@uantwerp.be}
\affiliation{Departement Fysica, Universiteit Antwerpen,
Groenenborgerlaan 171, B-2020 Antwerpen, Belgium}

\author{M. V. Milo\v{s}evi\'{c}}
\affiliation{Departement Fysica, Universiteit Antwerpen,
Groenenborgerlaan 171, B-2020 Antwerpen, Belgium}

\author{F. M. Peeters}
\affiliation{Departement Fysica, Universiteit Antwerpen,
Groenenborgerlaan 171, B-2020 Antwerpen, Belgium}
\date{\today}

\begin{abstract}
Magnetic flux patterns are known to strongly differ in the
intermediate state of type-I and type-II superconductors. Using a
type-I/type-II bilayer we demonstrate hybridization of these flux
phases into a plethora of unique new ones. Owing to a complicated
multi-body interaction between individual fluxoids, many different
intriguing patterns are possible under applied magnetic field, such
as few-vortex clusters, vortex chains, mazes or labyrinthal
structures resembling the phenomena readily encountered in soft
matter physics. However, in our system the patterns are tunable by
sample parameters, magnetic field, current and temperature, which
reveals transitions from short-range clustering to long-range
ordered phases such as parallel chains, gels, glasses and
crystalline vortex lattices, or phases where lamellar type-I flux
domains in one layer serve as a bedding potential for type-II
vortices in the other - configurations clearly beyond the
soft-matter analogy.
\end{abstract}
 \maketitle

\section{Introduction} Soft matter physics deals with systems as different as colloids,
polymers, gels, glasses, liquid crystals and others, where one
common feature is their self-organization into very rich mesoscopic
phases.\cite{Jones-book} To model this behavior, one often uses a pairwise inter-particle
interaction possessing several length scales and/or mixture of
attraction and repulsion.\cite{Malescio,reich,Camp,soft,klix,Dobnikar} Such interaction
potential, as a function of the particle density, indeed leads to
the formation of clusters, particle chains, labyrinthal gel-like
structures and (almost) regular lattices. This in turn questions the
known analogy between charged colloids and vortices in
superconductors, since the latter typically repel and form a
triangular (Abrikosov) lattice. On the other hand, type-I
superconductors are known to exhibit lamellar and labyrinthal flux
patterns, which lose distinction of individual vortices but resemble
the soft-matter structures in their macroscopic shape.\cite{Prozorov}
\begin{figure}[t]
\includegraphics[width=0.85\linewidth]{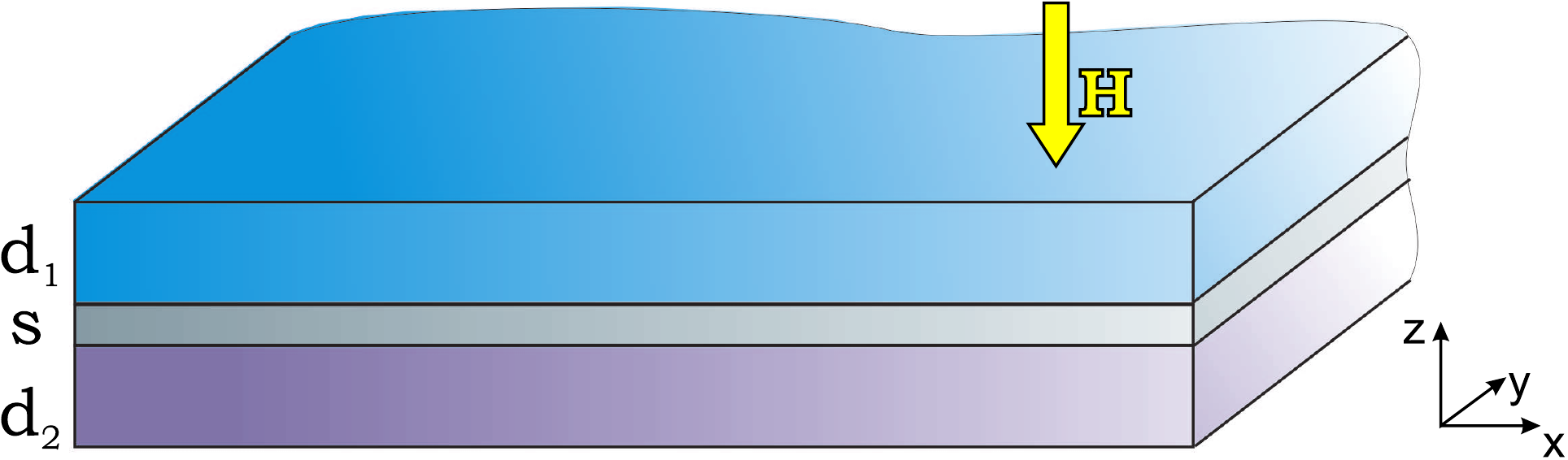}
\caption{The oblique view of the considered bilayer sample. The two
superconducting layers are separated by an ultrathin
oxide/insulating layer. The magnetic field is applied in the
direction perpendicular to the layers (along $z$-axis).}
\label{fig1}
\end{figure}

With this in mind, we here investigate magnetic flux patterns in a
coupled bilayer of two superconducting films - one type-I and one
type-II, under perpendicular magnetic field (see Fig.~\ref{fig1}),
in attempt to reveal unique vortex phases. In addition to the
crystalline vortex lattice, one now envisages vortex flocculation,
gelation and glassy phases, some similar to vortex matter
encountered in high-temperature,\cite{Blatter} multiband,\cite{VVM} and other unconventional superconducting\cite{Sigrist}
and superfluid systems.\cite{Eltsov} The film geometry is chosen
for an easy realization in experiment, but also in order to have
 asymptotic long-range $1/r$ repulsion between vortices\cite{Brandt} - similar to
the electrostatic Coulomb interaction in charged
colloids. We will show that the complexity of the obtained patterns
stems from the changes in the short- and mid-range interaction between vortices,
whose relative strength depends on the parameters of the layers,
especially their coherence length $\xi$ and penetration depth
$\lambda$, but also their thicknesses, electronic coupling between
them, and chosen temperature with respect to their individual
critical temperatures.

The paper is organized as follows. In Section II, we present the theoretical formalism. Section III summons and classifies the observed magnetic flux patterns, which are further characterized using radial distribution function in Section IV. Further we discuss the influence of temperature in Section V, where we also show the behavior of the heat capacity and its changes affiliated with different flux phases. Our results are summarized in Section VI. 

\section{Theoretical formalism} Most of earlier works on vortex structures and their dynamics employed molecular dynamics
with pairwise vortex interactions. This is truly valid only for
vortices in extreme type-II superconductors, where vortex cores are
point-like small. However, overlapping vortex cores do not interact
pairwise, and the interaction potentials are highly non-trivial.\cite{Andrey} This turns out to be even more complex for our
bilayer system, where vortices are extended objects with
different size of the core in two layers. We therefore opt for full
numerical simulation within the Ginzburg-Landau (GL) theory,
supplemented by the Lawrence-Doniach (LD) coupling between the
layers.\cite{LD,Klemm,Bulaevskii,Du,Graf} The appropriate free
energy functional then consists of the individual contributions from
each layer, the LD coupling term, and the energy of the magnetic
field in and around the sample:
\begin{widetext}
\begin{equation} \label{LDfunctional} \mathcal{F} =
\sum_{j=1,2} d_j \int \left[ \alpha_j |\Psi_j|^2 + \frac{1}{2}
\beta_j |{\Psi_j}|^4 + \frac{1}{2m_j}\left|\left(
\frac{\hbar}{i}\nabla-\frac{2e}{c}\mathbf{A}\right)
\Psi_j\right|^2\right] dS + s \int \eta|p\Psi_2 -\Psi_1|^2 dS + \int
\frac{(\mathbf h-\mathbf H)^2}{8\pi} dV.
\end{equation}
\end{widetext}
Here Cooper-pair condensates in the two layers are indexed by $j=1,2$ and
described by the order parameters $\Psi_j(\mathbf{r})$, assumed to
be uniform over the layer thickness $d_j$. {\bf H} is the applied
magnetic field and $\mathbf h = \mathrm {curl} \; \mathbf{A}$ is the
field including the magnetic response of the layers. The coefficients
$\alpha_{j}(T) = -\alpha_{j0}(1-T/T_{cj})$ are temperature dependent
(where $T_{cj}$ are the critical temperatures of the individual layers)
and connected with the nominal coherence lengths of the layers $\xi_j(T) =
{\hbar}/{|2m_j\alpha_j(T)|^{1/2}}=\xi_{j0}/\sqrt{1-T/T_{cj}}$, where
$m_j$ denotes Cooper-pair masses in the layers. The LD coupling
coefficient is $\eta = {\hbar^2}/(2 m_{\perp}s^2)$, where $m_{\perp}$
is the effective Cooper pair mass for tunneling between the layers
 and $s$ is the vertical distance between the layers (see Fig.~\ref{fig1}). 
The phase factor $p = \exp(- i \frac{2e} {\hbar c} \int_0^s A_z
dz)$ ensures the gauge invariance.

In what follows, we work in the London gauge $\nabla \cdot {\mathbf A} = 0$, therefore the Maxwell equation is just 
\begin{equation}
\label{maxwell}
-\triangle {\mathbf A} = \frac{4\pi}{c}{\mathbf j}. 
\end{equation}
Further we make approximation $A_z=0$, so that the phase factor $p$ in the LD coupling term is unity (similar model was also employed in Ref.~\onlinecite{Bluhm}). Since $A_z = 0$, Eq.~\eqref{maxwell} implies also $j_z = 0$. On the other hand, in the full model the current between the layers is 
\begin{align}
j_z &= \frac{i e \hbar} {m_\perp s} \left [ \psi_1^* p \psi_2  - p^* \psi_2^* \psi_1  \right ] \nonumber \\
 &=  \frac{2 e \hbar} {m_\perp s} |\psi_1| |\psi_2| \sin(\varphi_1 - \varphi_2), 
\end{align}
 where we used $p = 1$ and $\psi_j = |\psi_j| e^{i\varphi_j}$. 
The current between the layers $j_z$ can be neglected if it is much smaller than the currents within the superconducting layers 
\begin{eqnarray}
\label{LDequation_curr} \mathbf{j}_{j} = - \frac{ie\hbar}{m_j}
(\Psi_j^* \nabla \Psi_j - \Psi_j \nabla \Psi_j^*) - \frac{4e^2}{m_j
c}|\Psi_j^2|\mathbf{A}.
\end{eqnarray}
which leads to the condition
\begin{equation}
\frac{m_j}{m_{\perp}} \frac{\xi_j(T)}{s} \sin(\phi_1-\phi_2) \ll 1,
\end{equation}
if we assume that the amplitudes of the wave functions in two layers are not too different from each other. 
Assuming $A_z = 0$ is thus well justified for $m_{\perp} \gg m_j$ or very distant layers $s\gg \xi_j(T)$ i.e. in the case of very weakly coupled layers, but also in the opposite case $m_{\perp} \approx m_j$ where the phases of the two order parameters are almost the same, so that the term $\sin(\phi_1-\phi_2)$ vanishes. Therefore, for the parameters of the sample chosen in the present work $j_z$ is always much smaller than the current within the superconducting layers and one can safely neglect it.  

The variational minimization of the functional \eqref{LDfunctional}
(with $p = 1$) with respect to $\Psi_j^*$ leads to the GLLD equations
\begin{subequations}
\begin{equation}
\frac{1}{2m_1} \left( \frac{\hbar}{i} \nabla - \frac{2e}{c} \mathbf{A} \right)^2 \Psi_1 + \tilde{\alpha}_1 \Psi_1 +\beta_1 |\Psi_1|^2 \Psi_1 -\eta\frac{s}{d_1} \Psi_2 = 0,
\end{equation}
\begin{equation}
\frac{1}{2m_2} \left( \frac{\hbar}{i} \nabla - \frac{2e}{c} \mathbf{A} \right)^2 \Psi_2 + \tilde{\alpha}_2 \Psi_2 +\beta_2 |\Psi_2|^2 \Psi_2 -\eta\frac{s}{d_2} \Psi_1 = 0,
\end{equation}
\label{LDequations_dim}
\end{subequations} where $\tilde{\alpha}_j = \alpha_j + \eta s /d_j$. We solve
this system of equations numerically, self-consistently with the
equations for the supercurrent density per unit volume in each layer given by Eq.~\eqref{LDequation_curr}. Note that the supercurrents only flow inside the respective layers and therefore are spatially separated. Substituting Eq.~\eqref{LDequation_curr} into Eq.~\eqref{maxwell} with 
${\mathbf j}={\mathbf j}_1+{\mathbf j}_2$ provides the total
3D vector potential $\mathbf {A}$, which we calculated in the middle
of each layer and used iteratively in respective
Eqs.~(\ref{LDequations_dim}a,b) to compute $\Psi_1$ and $\Psi_2$
(for details of the procedure, see Ref.~\onlinecite{roel}).

We performed the numerical simulations on a rectangular region, with
aspect ratio 2 : $\sqrt{3}$ and periodic boundary conditions. We chose the parameter values close to a clean Nb
film as the type-II layer (with GL parameter $\kappa_1$
= 1.03, $T_{c1} = 9.2$ K) and Sn as the type-I layer ($\kappa_2 =
0.15$, $T_{c2} = 3.7$ K). The lowest considered temperature was 1K,
which was necessary to be deep in the superconducting state of the
type-I layer. Zero temperature coherence length of
the type-II layer ($\xi_{10}$) was then taken as unit of distance in all
calculations, while $\xi_{20}$ was swept between 2 and 10
$\xi_{10}$. Since this variation had only minor qualitative
influence on the observed vortex structures, we fixed the parameter
$\zeta=(\xi_{10}/\xi_{20})^2$ to 0.2, for computational convenience.
Order parameters were scaled to $\Psi_{j0} =
\sqrt{-\alpha_j(0)/\beta_j}$, and magnetic field to $H_0 = \Phi_0 /
(2\pi \xi_{10}^2)$, where $\Phi_0$ is the flux quantum.

\begin{figure}
\includegraphics[width=0.9\linewidth]{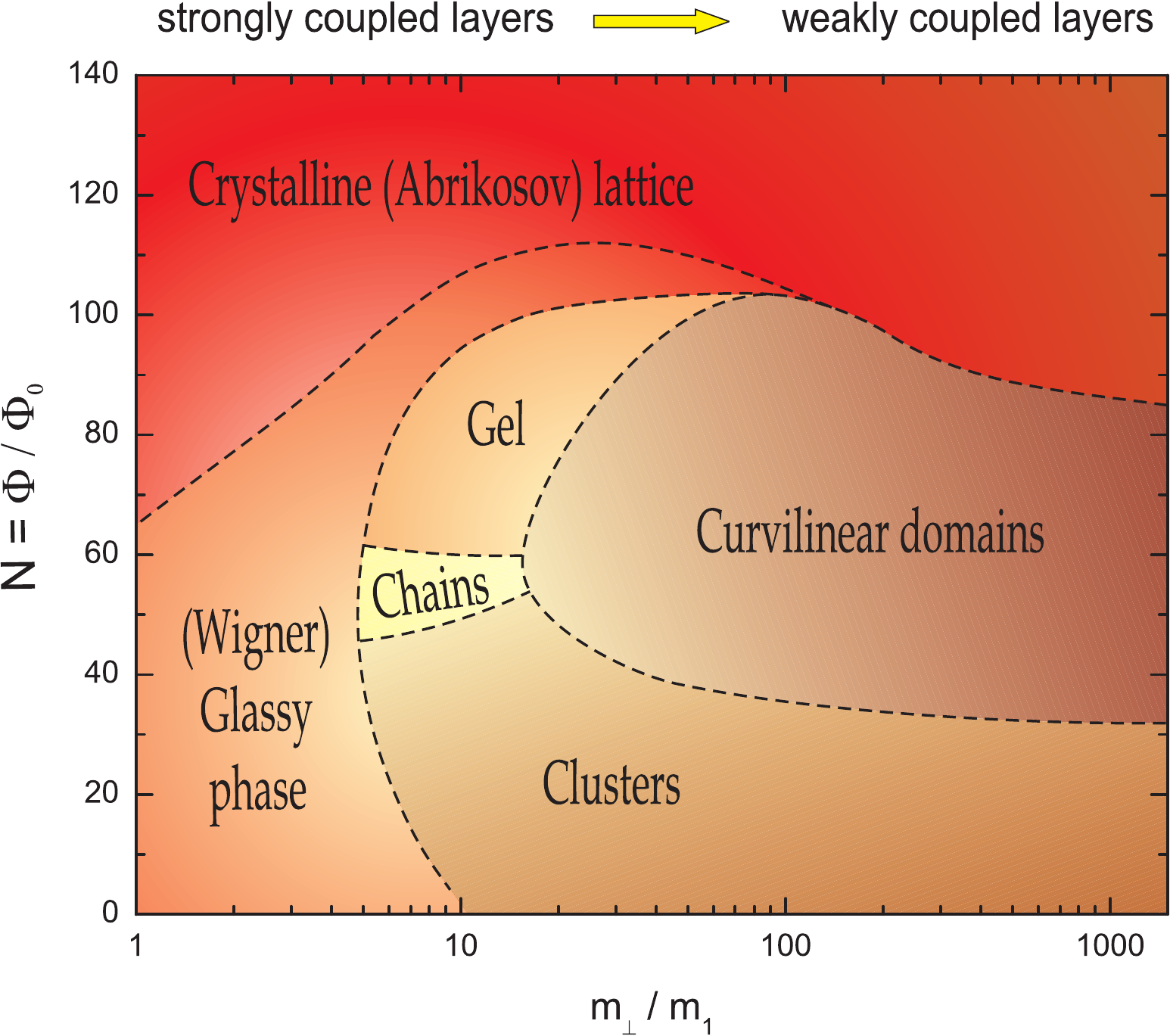}
\caption{The equilibrium phase diagram of a Nb/Sn bilayer calculated
at $T = 1$ K, for both layers $5\xi_{10}$ thick and spacer layer of
$0.05\xi_{10}$ in between, as a function of the applied field
(expressed through the number of flux quanta $N$ in the simulation
region 55 $\times$ 47.6 $\xi_{10}^2$) and effective mass $m_{\perp}$
of the Cooper pairs in the spacer layer. When other parameters are
fixed, the electronic coupling between the superconducting layers is
inversely proportional to $m_{\perp}$.} \label{fig2}
\end{figure}
\section{Phase diagram as a function of applied magnetic field and coupling between the layers} In Fig.~\ref{fig2}, we show the key
result of our paper - the vortex phase diagram of the described
bilayer at $T = 1$ K as a function of the applied magnetic field and
the effective mass in the spacer layer - which controls the strength
of the coupling between the superconducting layers. The field is
expressed through the number of vortices in the simulation region.
We revealed a series of nonuniform vortex phases, with phase transitions
between them indicated by dashed curves in the diagram. In what follows, we discuss
these transitions by showing exemplary vortex configurations.

\begin{figure}
\includegraphics[width=0.92\linewidth]{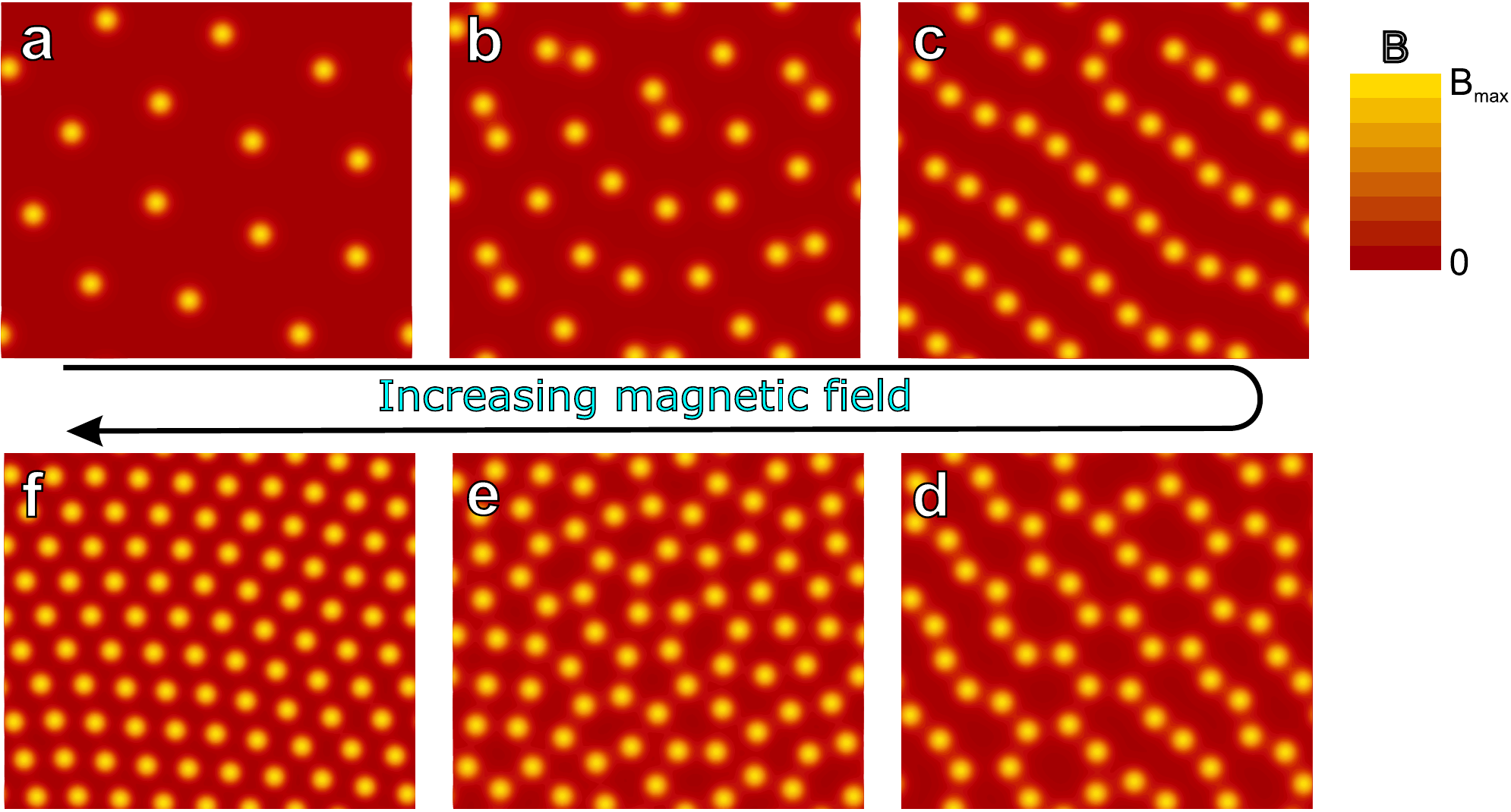}
\caption{{\it Very strongly coupled layers.} Vortex structure shown
via normalized magnetic field profile in the type-II layer
(identical vortex structure is found in the type-I layer),
corresponding to Fig. \ref{fig2} for $m_{\perp} = 5 m_1$. In panels
(a-f) there are 16, 32, 48, 64, 80 and 96 vortices in the simulation
region, respectively.} \label{fig3}
\end{figure}
Fig.~\ref{fig3} shows obtained vortex configurations for very
strongly coupled layers ($m_{\perp}=5m_1$). At low fields we observe
a quasi-homogeneous distribution of vortices, indicative of
long-range repulsion between them, reminiscent of Wigner glass in
soft-matter physics (see e.g. Ref.~\onlinecite{klix}). However, with
increasing field, after reaching some threshold vortex density, vortex dimers, short individual chains, and then long-ordered parallel chains are formed, in complete analogy to what was
seen in Ref.~\onlinecite{Malescio} for particle systems with purely
repulsive interactions but two governing length scales [i.e. a
potential with very strong repulsion at short range (``hard core''),
flattening at mid-range (``soft core'') and then abruptly weaker
repulsion at long range]. The half-distance between the parallel
vortex chains in Fig. \ref{fig3}(c) then gives an estimate of the
``soft core". As the vortex density is
further increased, the chains interconnect into a glassy disordered
structure, and finally form the crystalline (Abrikosov) lattice. The
intervortex distance at the latter glass-solid transition then
provides an estimate of the ``hard core'' of the repulsive vortex
interaction.

\begin{figure}
\includegraphics[width=0.8\linewidth]{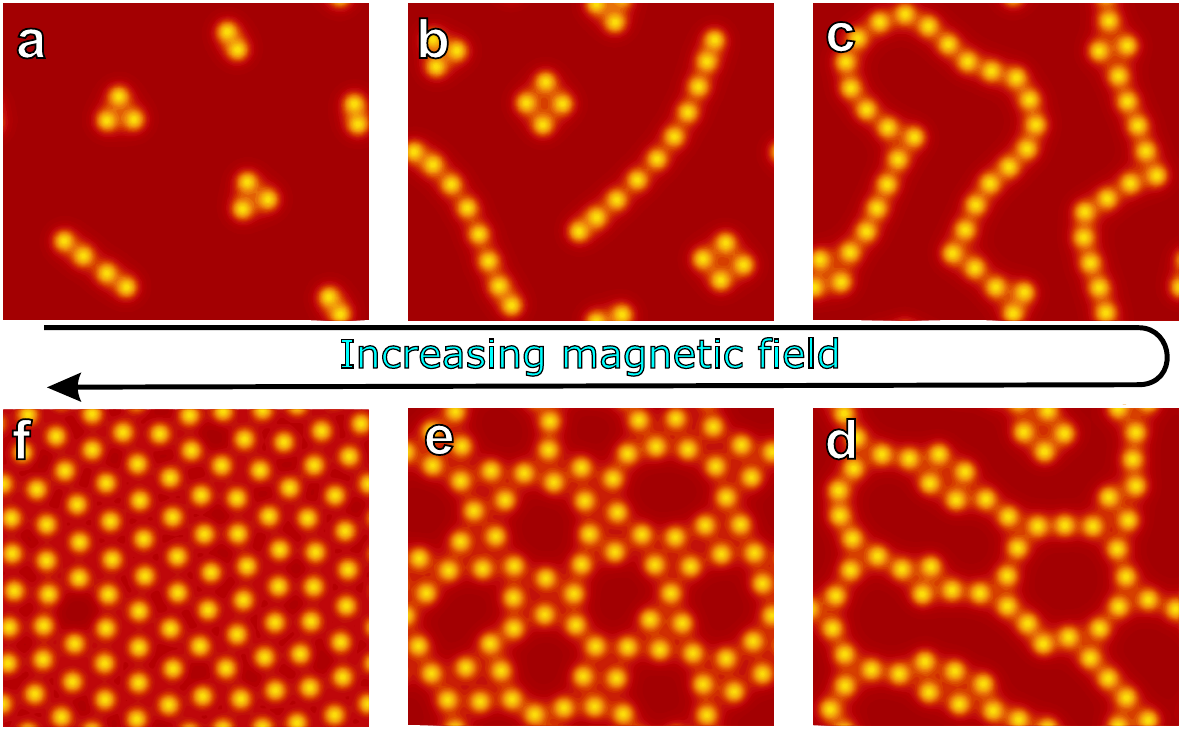}
\caption{{\it Strongly coupled layers.} Same as Fig. \ref{fig3}, but
for twice weaker coupling between the layers, i.e. $m_{\perp} = 10m_1$.
In panels (a-e) there are 16, 32, 48, 64, 80 and 96 vortices in the
simulation region, respectively.} \label{fig4}
\end{figure}
In Fig.~\ref{fig4}, we reduced the coupling between the layers with
a factor 2 as compared to Fig.~\ref{fig3} (i. e. we increased
$m_{\perp}$ to $10 m_1$). The found vortex configurations are very
different, starting at low fields from small clusters of 2-4
vortices, combined with short chains. With further increase of the
magnetic field the chains establish long-range order, then curve,
recombine, and finally interconnect into a low density network
filling the entire simulation region. Such a network, that spans the
volume of the medium while at low particle-density, is typical for
{\bf gels}. This gel-like structure is retained in Fig.~\ref{fig4}
until the newly added vortices fill all the voids in the gel and form
a disordered (glassy) lattice, similarly to the case of
Fig.~\ref{fig3}, followed by crystallization at high vortex density.

\begin{figure}
\centering
\includegraphics[width=0.8\linewidth]{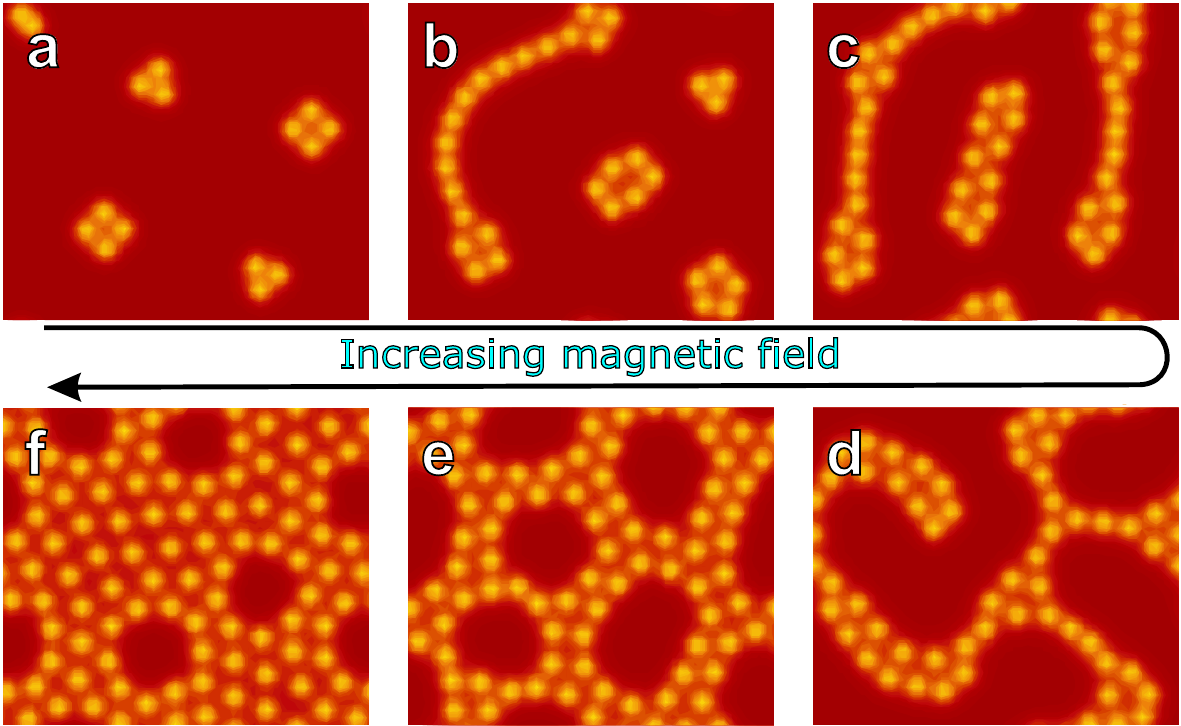}
\caption{Evolution of the states with applied magnetic flux, at coupling $m_\perp/m_1=15$ (c.f. Fig.~\ref{fig2}). The spatial profile of the magnetic field is shown for (a)-(f) 16, 32, 48, 56, 80 and 96 vortices respectively.
In this sequence of images, we sample the found phases in the busiest region of the phase diagram shown in Fig.~\ref{fig2}. Transitions between clusters and chains, to mazes and gel are shown.} \label{fig5}
\end{figure}
In Fig.~\ref{fig5}, the states are shown for $m_{\perp}=15 m_1$, which is still a relatively strong coupling. For low fields tiny clusters are formed, which then give way to prolonged chains, gel states and finally (quasi-)crystalline lattice for 128 vortices in the simulation box (not shown). One can see that $m_{\perp}=15 m_1$ lies on the crossover of stability regions of spatial structures with lateral extent one and two vortices, i.e. at the transition from chains to curvilinear domains in Fig.~\ref{fig2}. 
\begin{figure}
\includegraphics[width=0.8\linewidth]{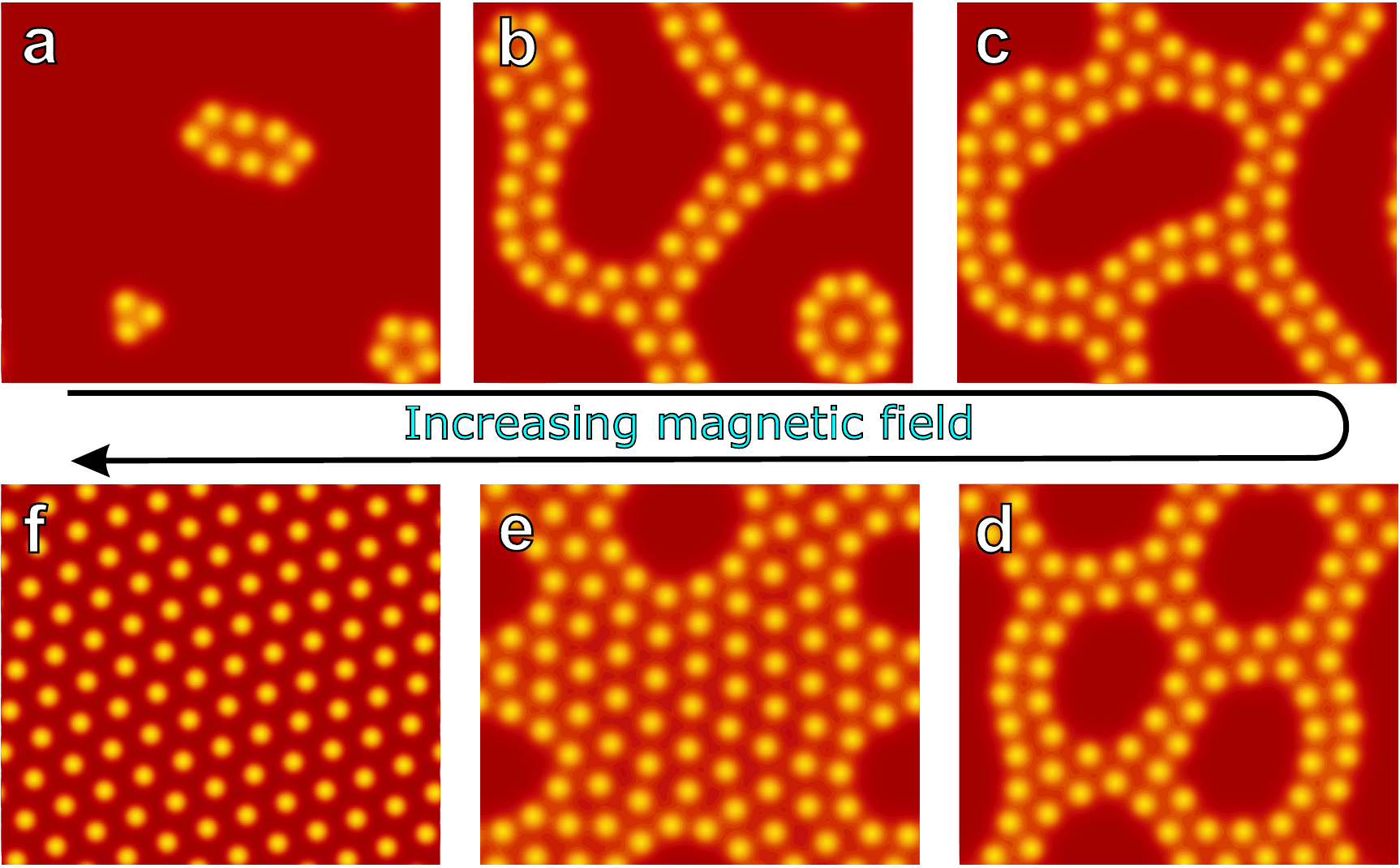}
\caption{{\it Intermediately strongly coupled layers.} Same as Figs.
\ref{fig3}-\ref{fig5} but for further weakened coupling between
layers, i.e. for $m_{\perp} = 20m_1$. In panels (a-f) there are 16,
64, 72, 80, 96 and 112 vortices in the simulation region,
respectively.} \label{fig6}
\end{figure}
\begin{figure}
\includegraphics[width=0.89\linewidth]{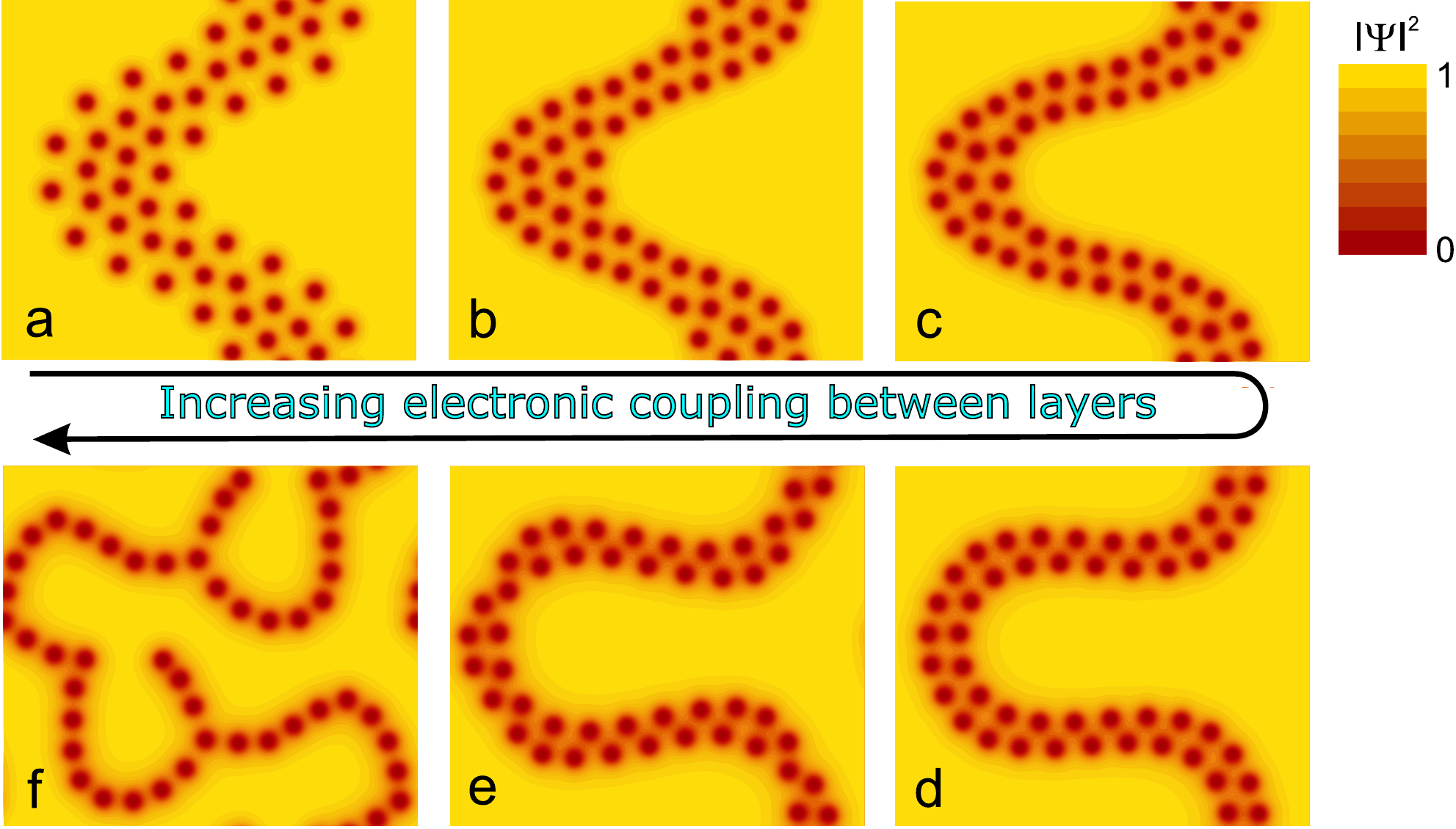}
\caption{The evolution of the vortex structure with decreasing
$m_{\perp}$, i. e. with increasing strength of the electronic coupling
between the layers, for 48 vortices in the simulation region. In panels
(a-f), the normalized Cooper-pair density in the type-II layer is
shown for $m_{\perp}/m_1 = 160$, $120$, $60$, $30$, $20$ and $10$,
respectively.} \label{fig7}
\end{figure}
In Fig. \ref{fig6}, the electronic coupling between superconducting
layers is further decreased ($m_{\perp}=20 m_1$), and the
qualitative trend of Fig. \ref{fig5} is maintained. With increasing
field, a transition from a solution of clusters to short-range mazes
and then long-range gel is found, followed by crystallization of the
vortex lattice. However, the superconducting type-I behavior of the
type-I layer becomes more apparent, as vortices become less distinct
from each other, and occupy increasingly wider domains. This is more
clearly demonstrated in Fig. \ref{fig7}, where we gradually
increased the coupling between the layers for a fixed number of
vortices in the simulation. Notice that the lateral extent of the
vortex stripes varies from just one vortex for strong coupling, to
four vortices for weak coupling. This behavior is reminiscent of the
colloidal structures studied in Ref.~\onlinecite{reich}, where particles
interacted via Coulomb-like repulsion combined with a mid-range
attraction, and where the widening of the stripe phases was directly
linked to the increasing strength and range of the attractive part
of the interaction.

\begin{figure}
\centering
\includegraphics[width=0.6\linewidth]{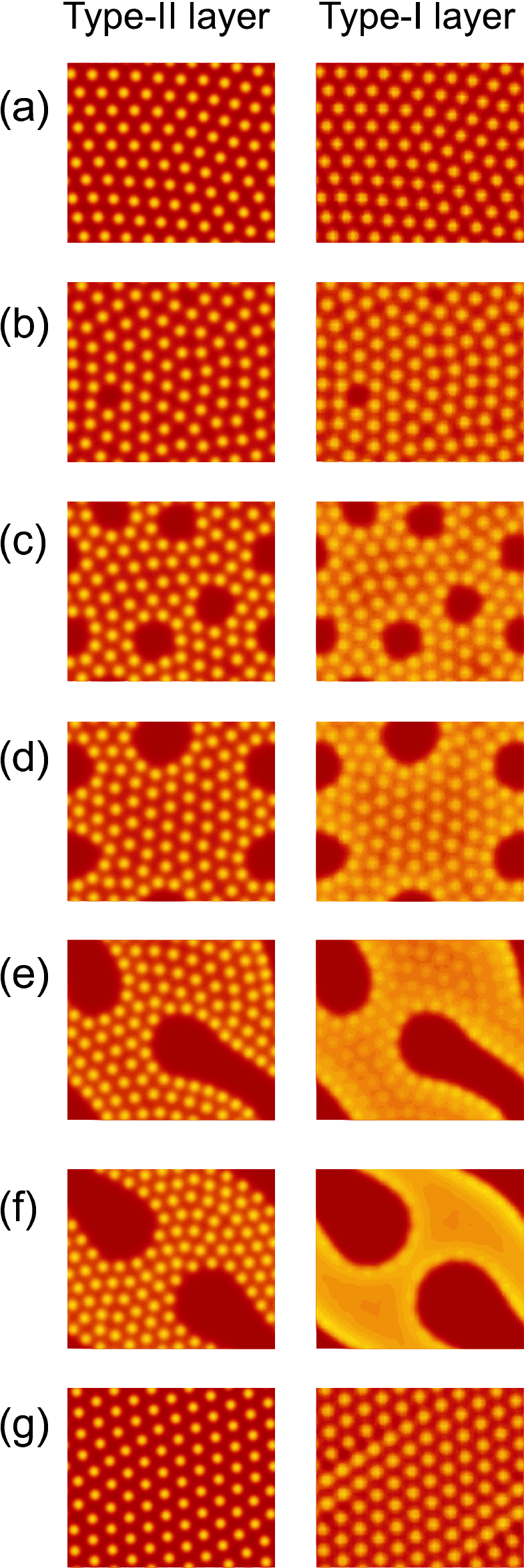}
\caption{Evolution of the vortex states with decreasing coupling, at fixed magnetic flux (96 vortices, c.f. Fig.~\ref{fig2}). The spatial profile of the magnetic field is shown. From top down $m_\perp/m_1 =$ (a) 5, (b) 10, (c) 15, (d) 20, (e) 40, (f) 200 and (g) 300. Left panel shows the type-II layer and the right one type-I layer. In this sequence of images, we show the found phases going left-to-right in the phase diagram shown in Fig.~\ref{fig2}. Transitions between Abrikosov lattice, gels, and type-I domains decorated by type-II vortices are found. Note the differences between the left (type-II) and the right (type-I) panel, demonstrating that for $m_\perp/m_1>20$ the notion of individual vortices is gradually lost.} \label{fig8}
\end{figure}
\begin{figure}
\includegraphics[width=0.9\linewidth]{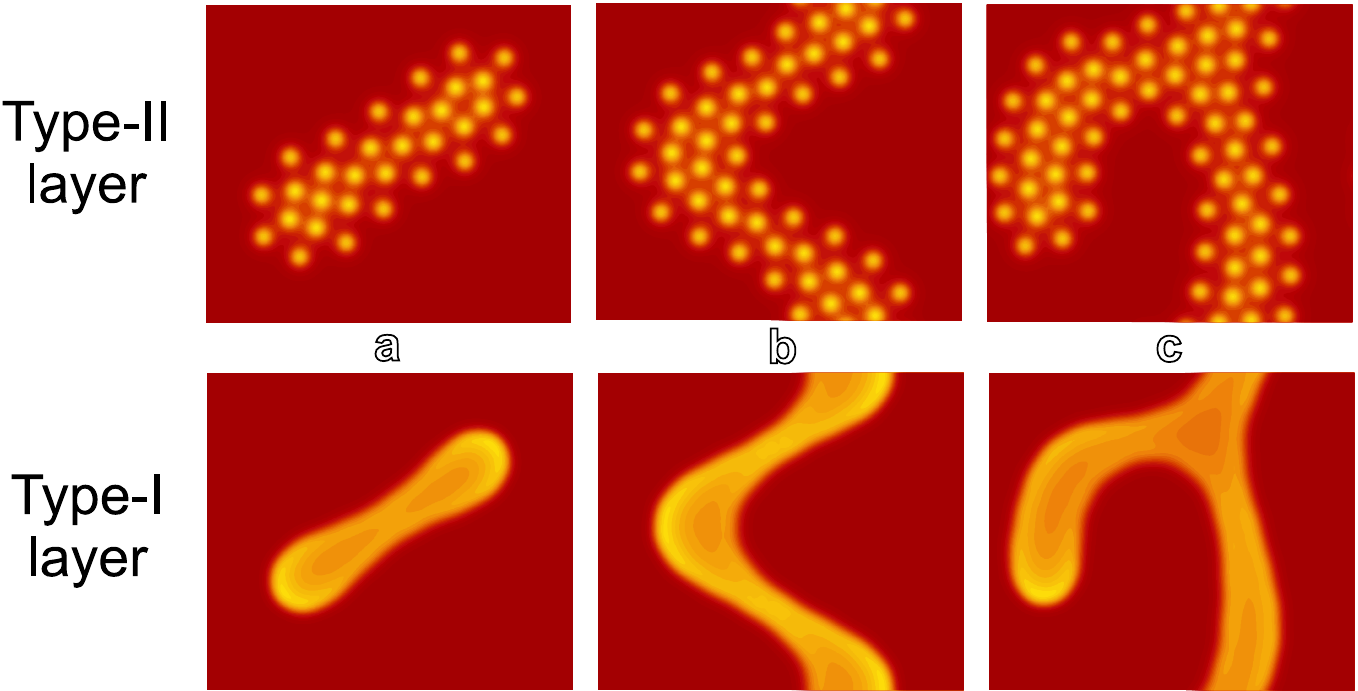}
\caption{{\it Weakly coupled layers.} Flux patterns shown in the
type-II layer (top row) and the type-I layer (bottom row),
corresponding to Fig. \ref{fig2} for $m_{\perp} = 1000m_1$. In
panels (a-c) there are 32, 48, and 64 flux quanta in the simulation
region, respectively.} \label{fig9}
\end{figure}
For even smaller values of interlayer coupling one can no longer rely on a two-body vortex-vortex interaction, since the type-I layer exhibits laminar domains, and the notion of individual vortices is completely lost. In Fig.~\ref{fig8} we depict the changes in vortex patterns in a broad range of $m_{\perp}$ values, where formation of type-I domains is visible for $m_{\perp} > 20m_1$. Those domains act as potential wells for vortices in the type-II layer, as exemplified in Fig.
\ref{fig9}, which shows states for $m_{\perp} = 1000m_1$. Because of this, the structural phase transitions become different from the $m_{\perp} < 20m_1$ cases. This is clearly seen in the change of curvature of the sol-gel and the gel-glass
transition lines in Fig. \ref{fig2}. With decreasing coupling, due
to the easier formation of type-I domains, vortices in type-II layer
connect into mazes at lower densities. On the other hand, they also
crystallize at lower densities than for strong coupling, which is
due to the practically destroyed superconductivity in the type-I
layer at such a large magnetic field. The formation of domains in
type-I layer as a trapping potential for vortices is interesting because of their sensitivity to applied in-plane magnetic field,\cite{wijng} or current.\cite{hoberg} Therefore one can easily manipulate externally the domain structure in the type-I
layer (e.g. straighten/relax the domains), and thereby dynamically
change and restore the vortex patterns in the type-II layer,
similarly to the controllability achieved in the
superconductor-ferromagnet bilayers.\cite{Vlasko-Vlasov}

\section{Radial distribution functions of observed configurations}
In this section we present the radial distribution functions $g(r)$ of different vortex phases observed in the superconducting type-I/type-II
bilayer, which can in some cases serve to distinguish particular vortex configurations in e.g. SANS measurements.

\begin{figure}
\includegraphics[width=\linewidth]{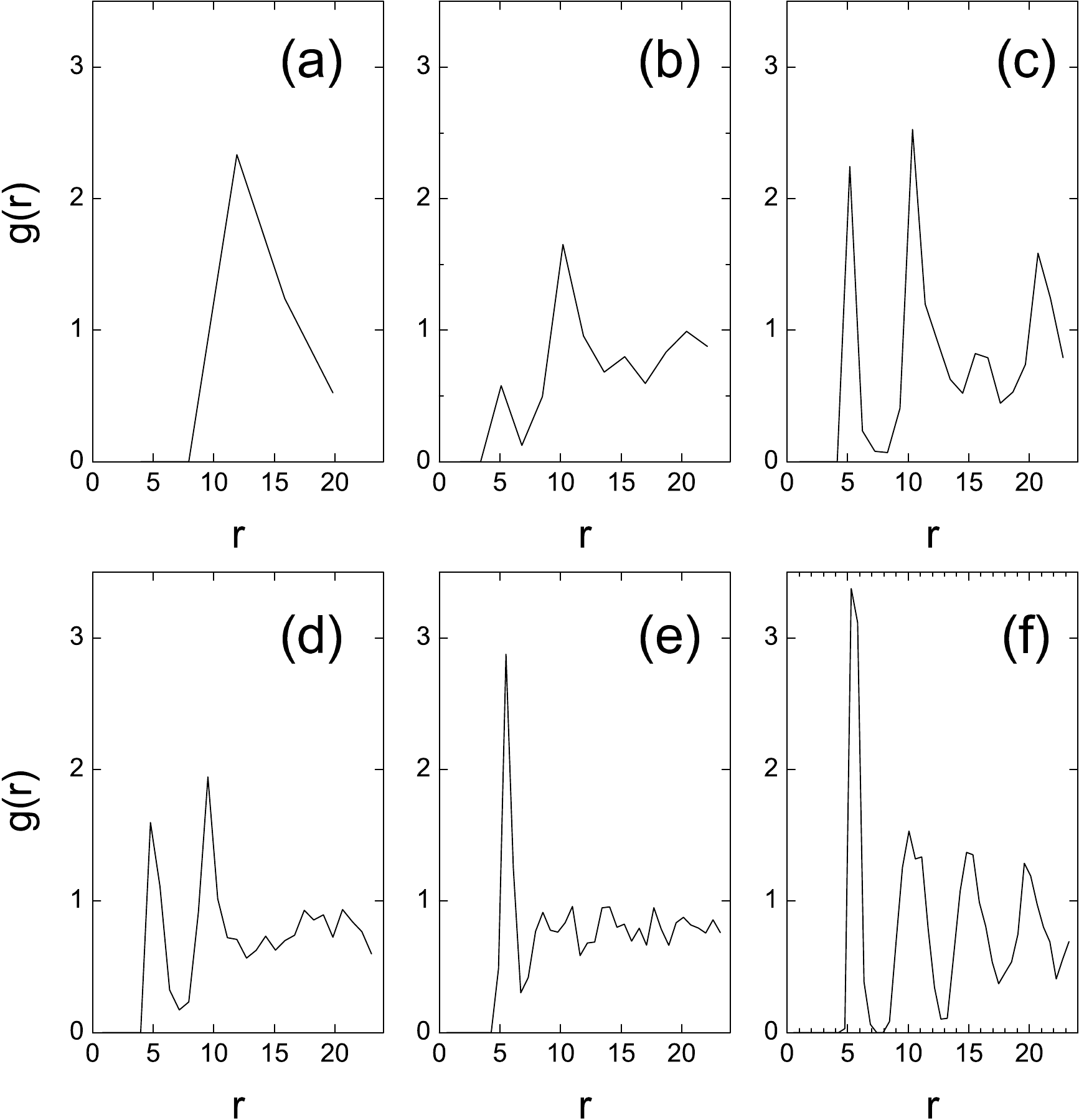}
\caption{The radial distribution function $g(r)$ for $m_{\perp} = 5
m_1$. Panels (a-f) correspond to states with 16, 32, 48, 64, 80 and
96 vortices in the simulation region, respectively.} \label{fig10}
\end{figure}

\begin{figure}
\includegraphics[width=\linewidth]{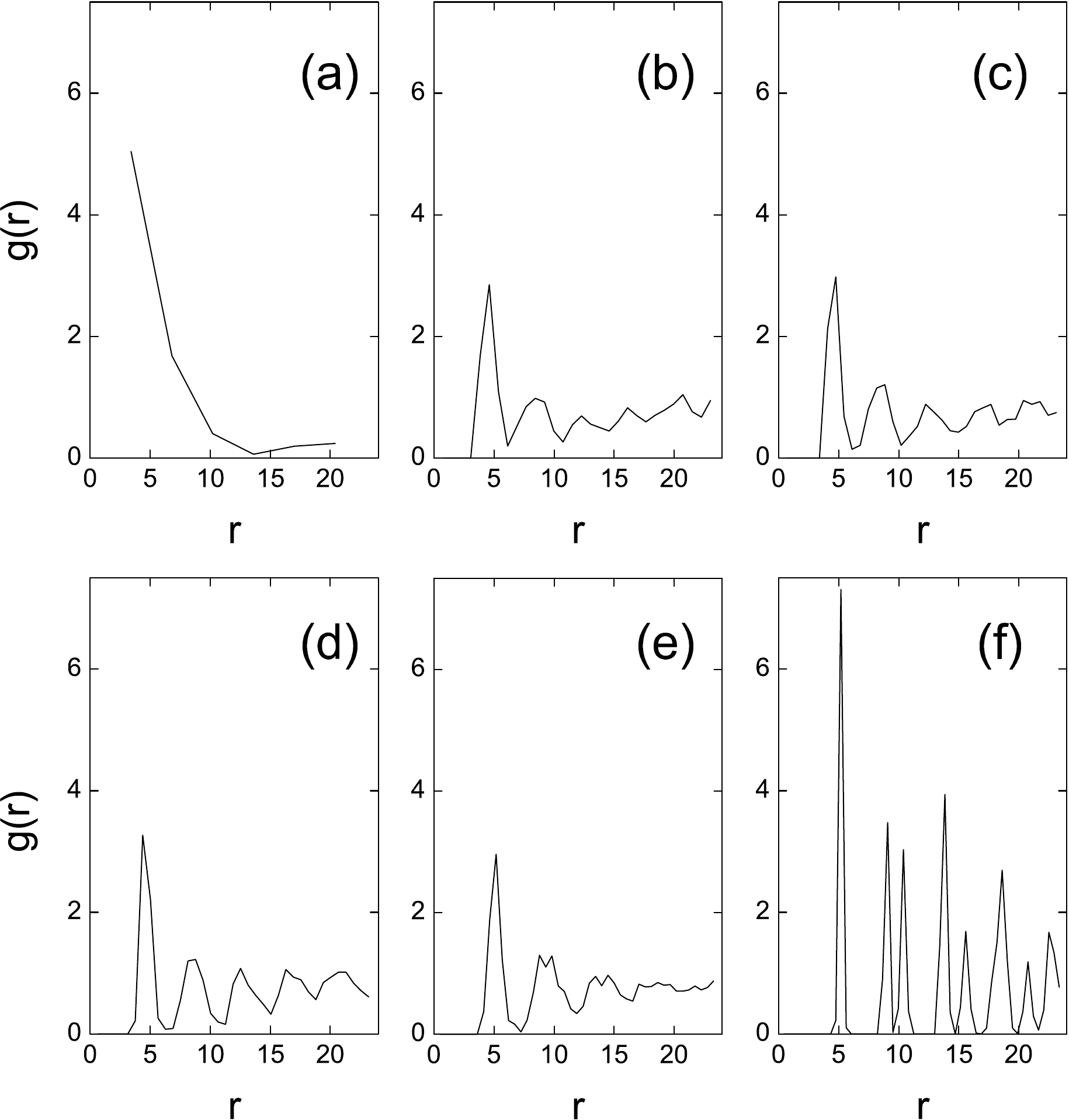}
\caption{The radial distribution function $g(r)$ for $m_{\perp} = 20
m_1$. Panels (a-f) correspond to states with 16, 64, 72, 80, 96 and
112 vortices in the simulation region, respectively.} \label{fig11}
\end{figure}

The radial distribution function $g(r)$ characterizes the particle distribution. It gives clear signatures of the crystalline order and can be also experimentally determined e.g.
by neutron scattering. It is defined as
\begin{equation}
g(r) = \frac{1}{N} \frac{\rho(r)}{\rho},
\end{equation}
where $\rho(r)$ is the density of particles at some distance $r$ from the
origin, while $\rho = N/A$ - with $N$ the total number of particles
and $A$ the total area - is the average density.
The density $\rho(r)$ is in practice computed from the histogram of
all distances between the pairs of particles, where the number of
particles $N_i$ in each bin $[r_i-dr/2;r_i+dr/2]$ is divided by the
area corresponding to that bin $A_i = \pi \left [(r_i + dr/2)^2 -
(r_i - dr/2)^2\right ]$. Note that each pair counts for two
particles. Taking this into account, we calculated $g(r)$ using the 
formula
\begin{equation}
g(r) = \frac{2N_i}{N \rho \pi \left [(r_i+dr/2)^2-(r_i-dr/2)^2 \right ]}.
\end{equation}
In the Figs.~\ref{fig10} and \ref{fig11} we show the radial distribution
function for the states displayed in Figs.~\ref{fig3} and \ref{fig6} respectively. In each case we use the total number of bins equal to half of the total number of vortices. The distance between any two vortices is
determined as the distance to the nearest periodic image. In order
not to account the interaction of vortex with its own periodic
image, we only show $g(r)$ for $r$ up to 23 $\xi_{10}$ i.e.
approximately half of the shorter side of our simulation region.

The $g(r)$ functions for smaller densities, i.e. for 16 and 32
vortices in the simulation region are too coarse to be useful,
however in the radial distribution functions for our simulations for
larger applied magnetic fields we can see important signatures of
the order. For example in Fig.~\ref{fig11} in panels (e) and (f) for 96, respectively  112 vortices one can see that the second peak splits into two, which is a well-known effect due to the formation of the regular triangular
(Abrikosov) lattice (in the first case with holes of the Meissner
state still present).
One can also see in Fig.~\ref{fig10} that the chain phase for $m_{\perp}/m_1 = 5$ and intermediate densities (48 and 64 vortices)
can be distinguished by having the second peak of $g(r)$ higher than
the first one. Similar signature was also observed in
Ref.~\onlinecite{Malescio}, but there it was found for the structure
factor $S(k)$ i.e. in the $k$-space. This signature
disappears when a gel state is formed.

\section{Phase transitions between different structures with temperature}
\begin{figure}
\includegraphics[width=0.8\linewidth]{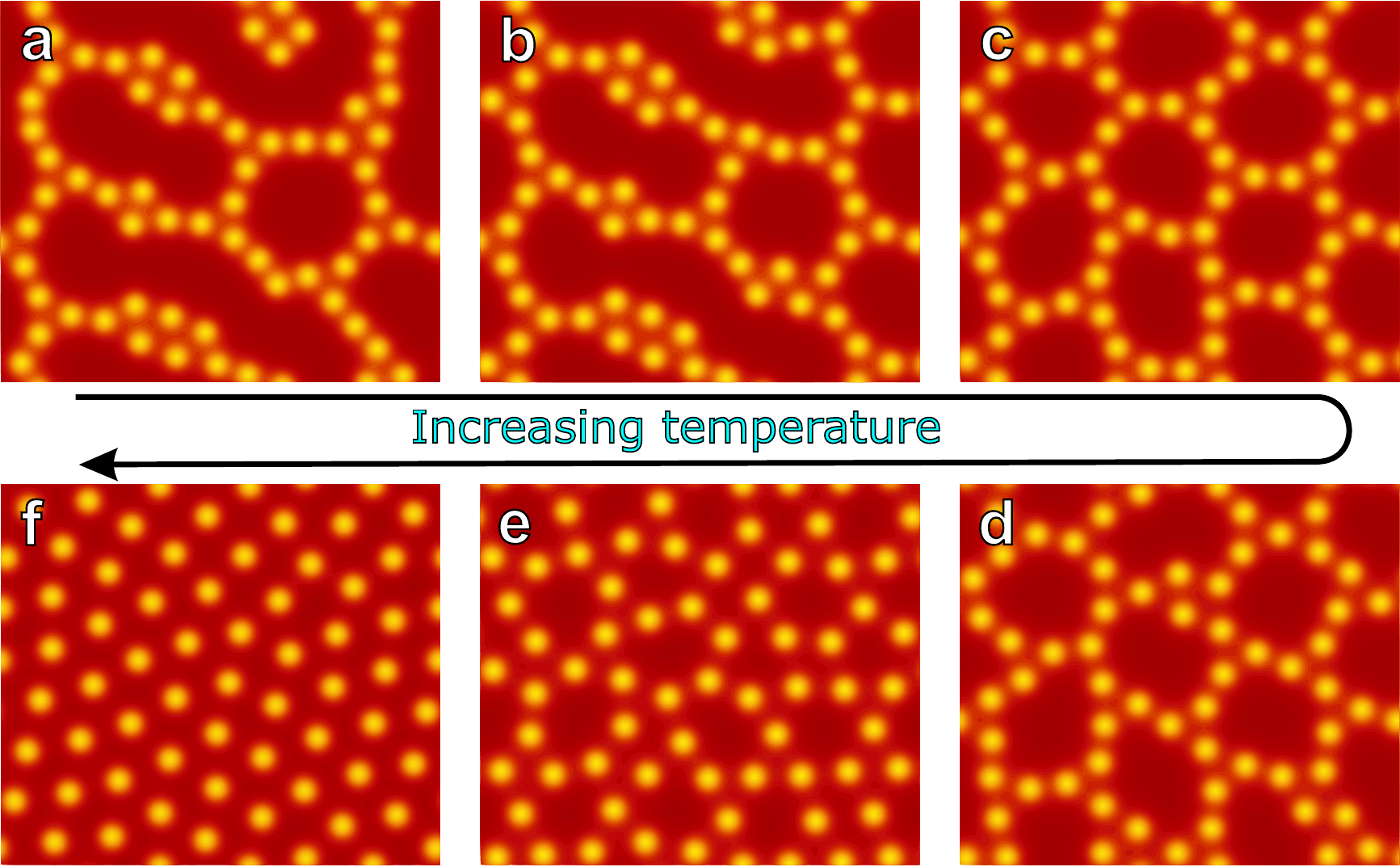}
\caption{The evolution of the gel-like phase of Fig. \ref{fig4}(d)
with increasing temperature. The vortex structure in the type-II
layer is shown for (a) $T = 1$ K (original state) and the
field-heated states (b) $T = 1.1$ K, (c) 1.25 K, (d) 1.45 K, (e)
1.85 K, and (f) 2.2 K.} \label{fig12}
\end{figure}

\begin{figure}
\centering
\includegraphics[width=0.8\linewidth]{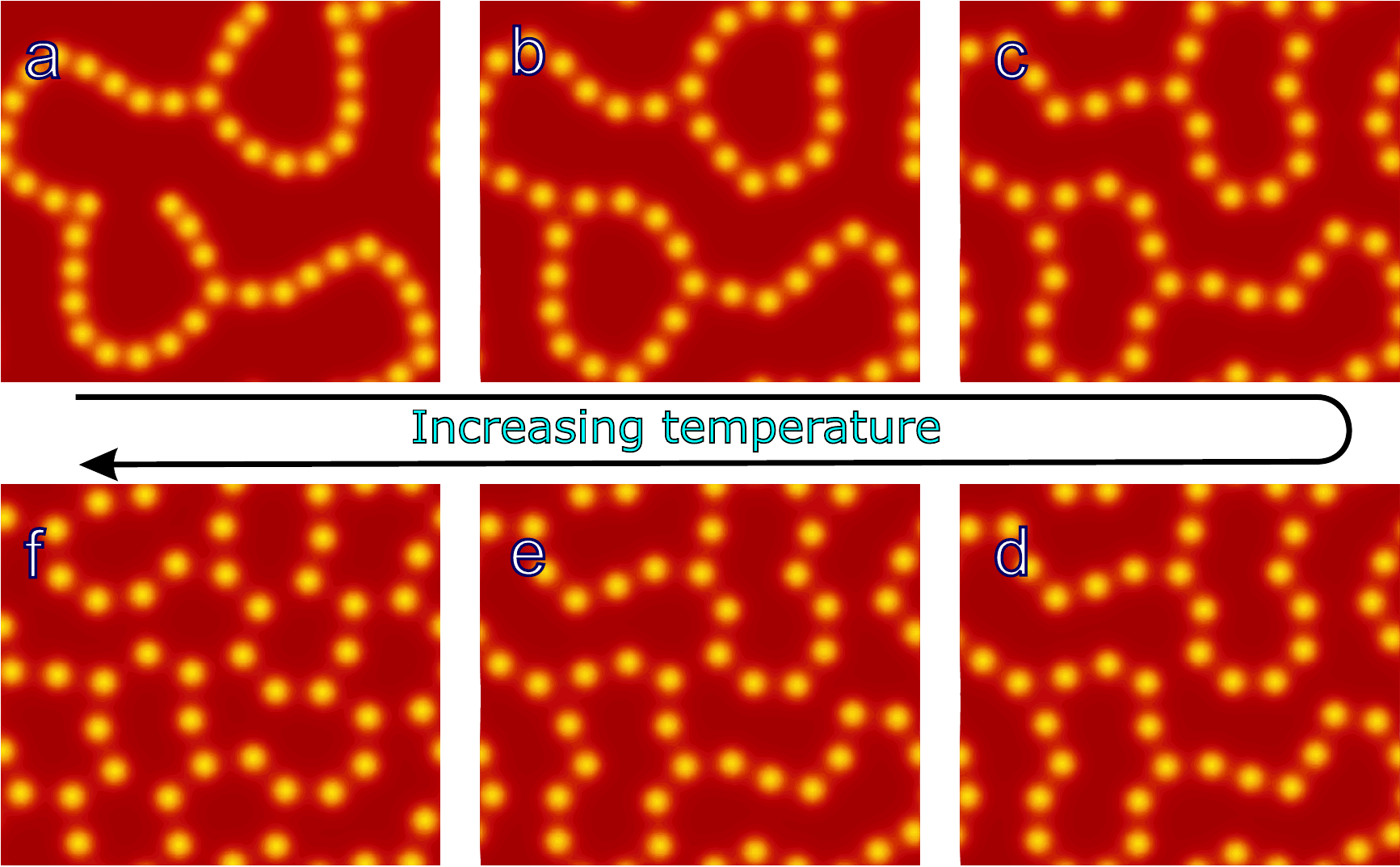}
\caption{Evolution of the state shown in Fig.~\ref{fig7}(f), with increasing temperature. The spatial profile of the magnetic field is shown.
In this sequence of images, we show the transition from the chain phase to the gel phase by gradual increase of temperature. (a)-(f) $T=$ 1 K, 1.5 K, 1.75 K, 1.85 K, 1.9 K and 2 K. } \label{fig13}
\end{figure}

Another degree of controllability of the flux patterns in our system is provided by
temperature, due to the different critical
temperatures $T_{cj}$ of the layers. For the considered parameters,
elevated temperature will swiftly deplete superconductivity in the
type-I layer, and will also interconnect the flux patterns in that
layer due to the increasing coherence length. Both these features
will influence the observed vortex configurations in the type-II
layer. Hence, one is able to control and monitor the phase
transitions of soft vortex matter in our bilayer system simply by
changing temperature. We illustrate this in Fig. \ref{fig12}, where
we start from a rather disordered gel-like state from Fig.
\ref{fig4}(d) and gradually increase temperature. After a transition
to a honeycomb (or fishing net) structure at $T = 1.25$ K, a
transition to a glassy phase was found at 1.8 K, followed by
crystallization into the Abrikosov lattice at 2.1 K (where the
magnetic influence of the type-I layer became negligible). It is therefore interesting to note that contrary to most natural structures, including soft matter, the vortex configurations in our system become {\it more ordered} with increasing temperatures. Similarly, in Fig.~\ref{fig13} one can see how increasing the temperature transforms the chain state of Fig.~\ref{fig7}(f) to the gel phase.

\begin{figure}
\includegraphics[width=0.8\linewidth]{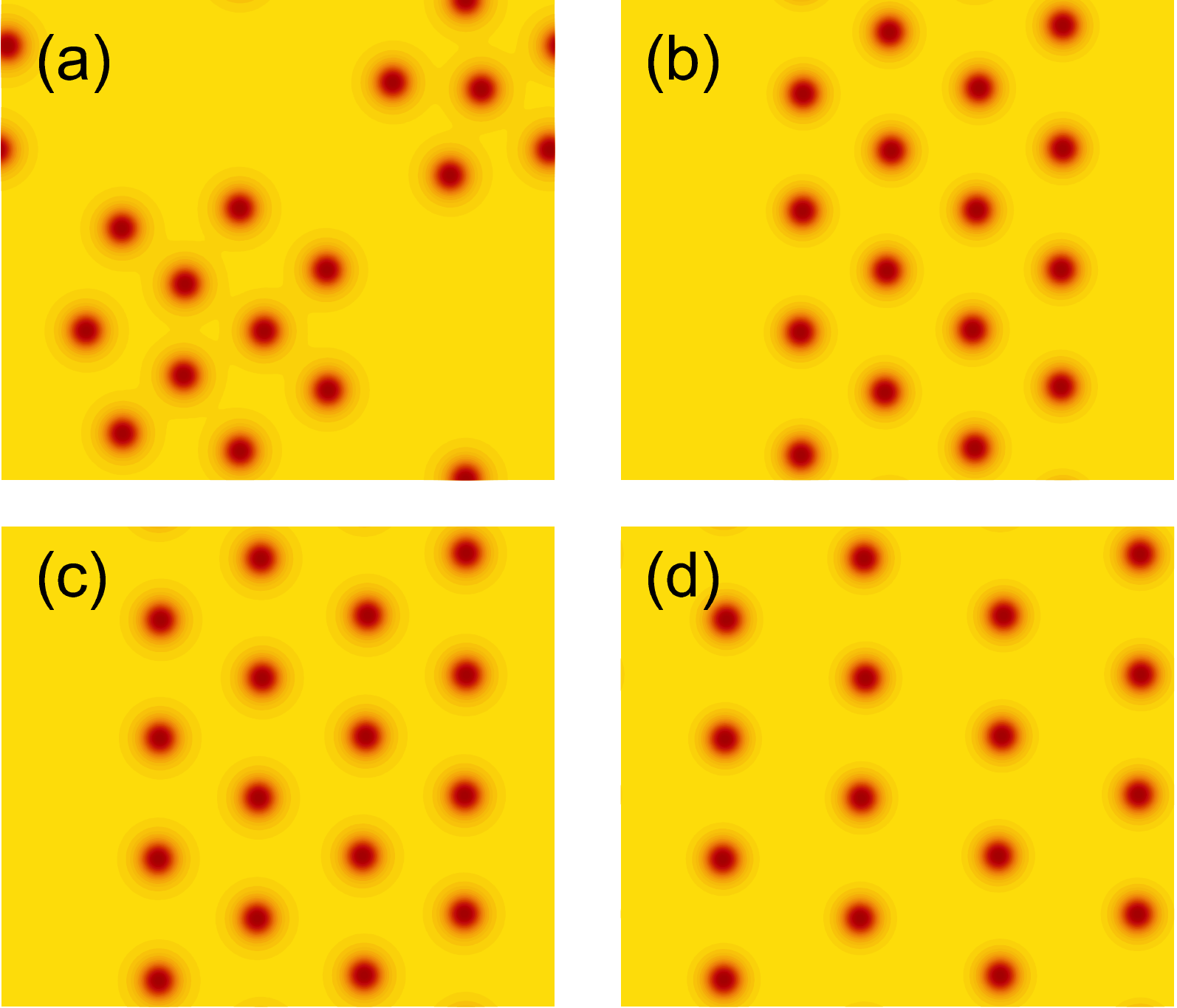}
\caption{The transition from the cluster phase through stripe phase
to the Abrikosov lattice \emph{on heating}, shown as Cooper-pair
density plots in the type-II layer at (a) $T = 3.1$ K (clusters),
(b) $T = 3.15$ K, (c) $T = 3.25$ K (stripe) and (d) $T = 3.3$ K
(Abrikosov lattice).} \label{fig14}
\end{figure}

\begin{figure}
\includegraphics[width=0.8\linewidth]{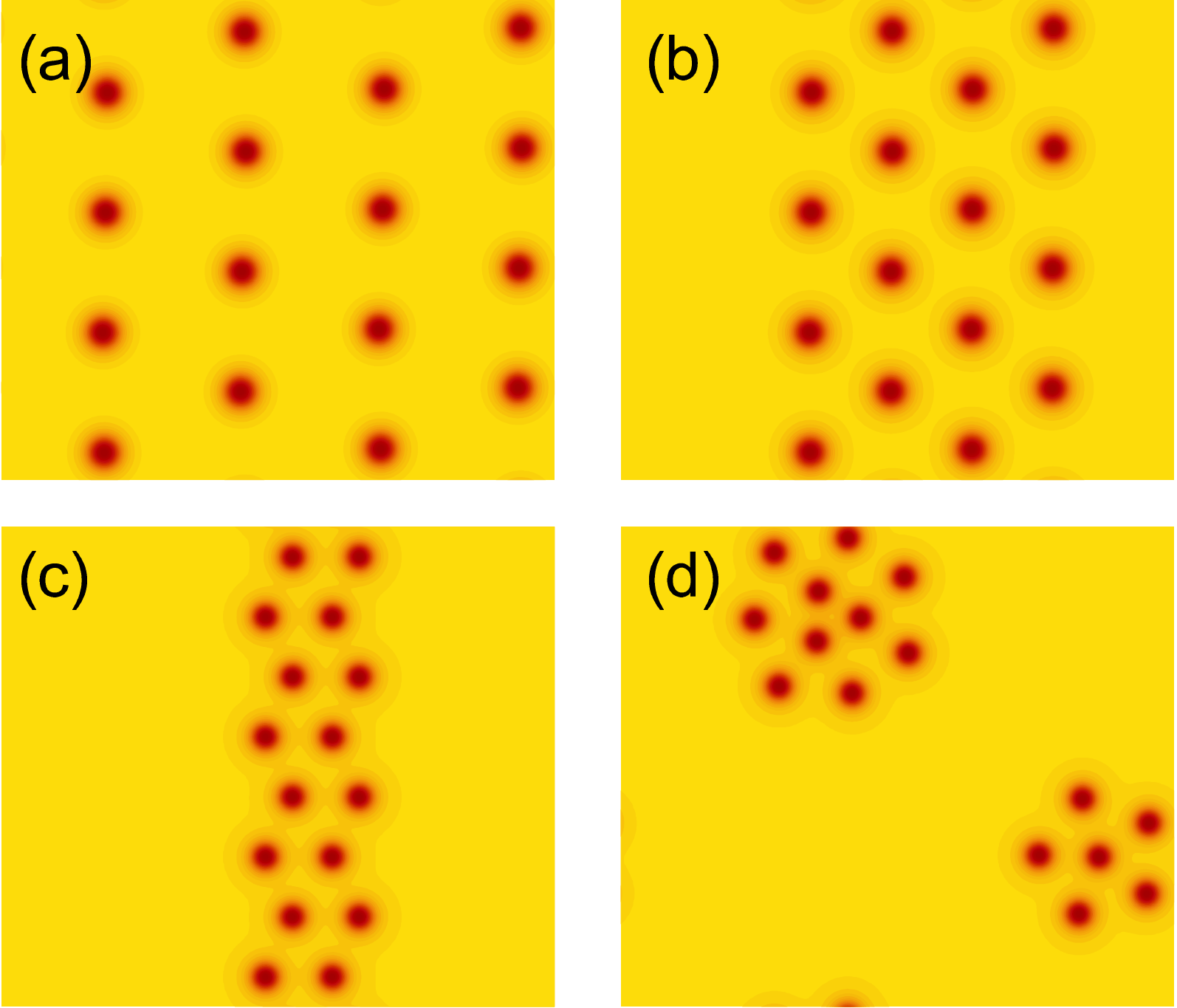}
\caption{The transition from the Abrikosov lattice through stripe
phase to the cluster phase \emph{on cooling}, shown as Cooper-pair
density plots in the type-II layer at (a) $T = 3.15$ K (Abrikosov
lattice), (b) $T = 3.1$ K, (c) $T = 0.85$ K (stripe) and (d) $T =
0.84$ K (clusters).} \label{fig15}
\end{figure}

To gain insight into the nature of the phase transitions between
different spatial arrangements of vortices, we show one particular
example calculated for 16 vortices, thickness of the coupling layer
$s = 2 \xi_{10}$ and effective mass for tunneling of the Cooper
pairs between the layers $m_\perp = 40 m_1$. We found that for
these parameters the stable phase at low temperature are small
vortex clusters. These for intermediate temperatures coalesce into a
single stripe which at higher temperature spreads over entire
simulation region and interconnects with adjacent stripes into
Abrikosov lattice. However, we found that there is considerable
hysteretic behavior, since Fig.~\ref{fig14} shows the transition from several clusters to a stripe between 3.1 K and 3.15 K with increasing temperature, while on cooling (Fig.~\ref{fig15}) the stripe is stable down to a much lower temperature of 0.85 K.
In order to characterize this phase transition, we calculate the
free energy and heat capacity. The free energy is given by Eq.~\eqref{LDfunctional} which in dimensionless form reads
\begin{align} &\frac{\mathcal{F}}{\mathcal{F}_0} =
 d_1 \int \left[ -\chi_1 |\Psi_1|^2 + \frac{1}{2} |{\Psi_1}|^4 + \left|\left(
-i\nabla-\mathbf{A}\right)
\Psi_1\right|^2\right] dS \nonumber \\ &+ C_1 d_2 \int \left[ -\chi_2 |\Psi_2|^2 + \frac{1}{2} |{\Psi_2}|^4 +  \frac{1}{\zeta}\left|\left(
-i\nabla-\mathbf{A}\right)
\Psi_2\right|^2\right] dS  \nonumber \\ &+ C_2 \int |\Psi_1-C_3\Psi_2|^2 dS + \kappa_{1}^2 \int
(\mathbf{A}-\mathbf{A}_0)\cdot \mathbf{j} dV,
\end{align}
where $\mathcal{F}_0 = \xi_{10}^3 \alpha_{10}^2/\beta_1 = \Phi_0^2 /(32\pi^3 \kappa_1^2 \xi_{10})$ is our unit
of energy ($\mathcal{F}_0 \approx 1.07 \times 10^{-18}$ J for Niobium with $\xi_{10}=$ 38 nm), $\chi_j = 1 - T/T_{cj}$, $\mathbf{A}_0$ is the vector potential of the applied
field, $\mathbf{j}$ the supercurrent, and $C_1 = \zeta^2 \frac{\kappa_1^2}{\kappa_2^2}$, $C_2 = \frac{m_1}{s m_\perp}$ and $C_3 = \frac{\kappa_1}{\kappa_2} \sqrt{\zeta\frac{m_2}{m_1}}
$. 

\begin{figure}
\includegraphics[width=0.95\linewidth]{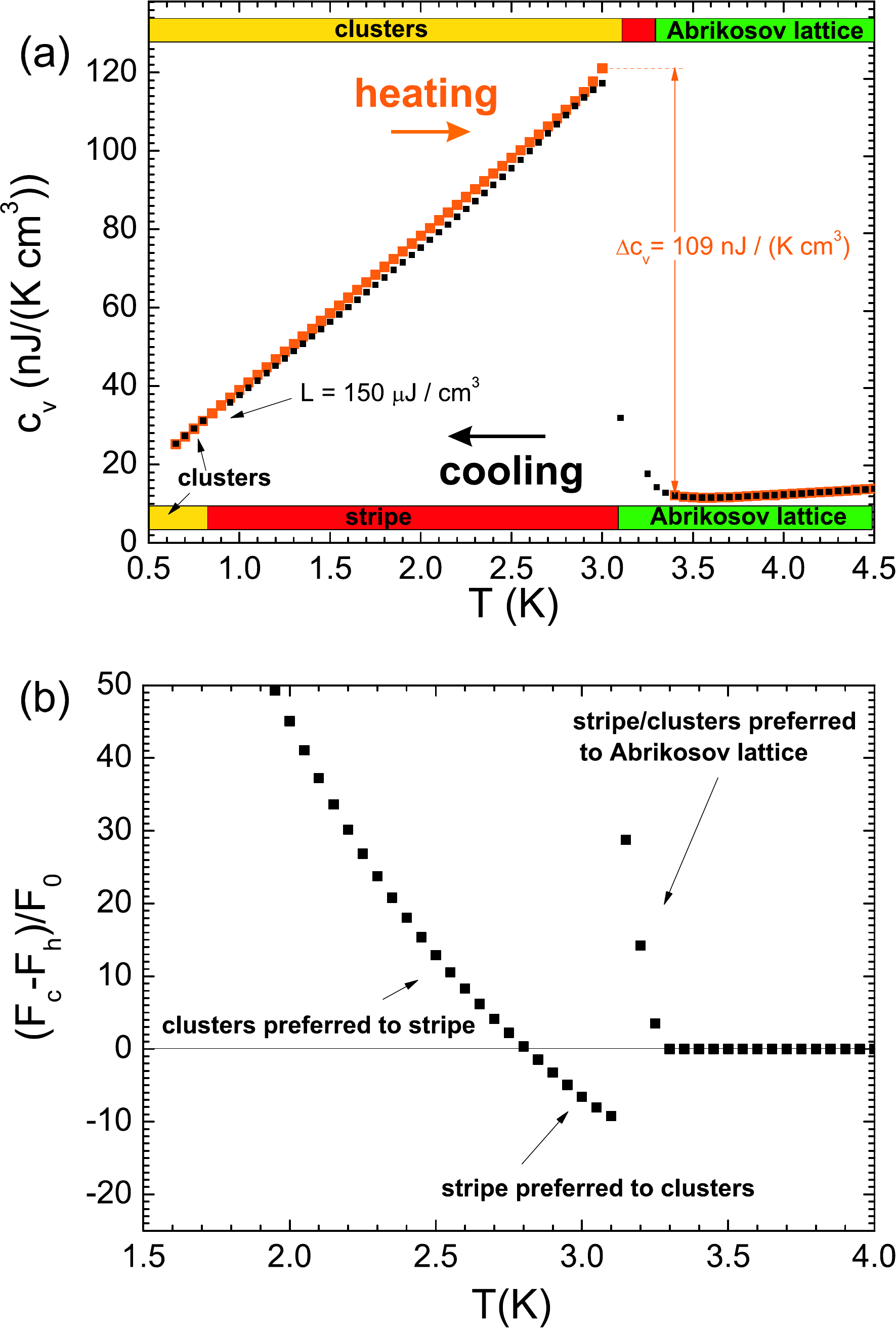}
\caption{(a) The hysteretic behavior between the cluster phase, the
stripe phase, and the Abrikosov lattice, shown via calculated heat
capacity $c_v$ on heating and cooling. The top (bottom) banner label the states found on heating (cooling). The major difference is that on heating the stripes are only found in a narrow temperature region close to $T =$ 3.1 K, while on cooling they are stable down to 0.85 K. (b) The difference in the
free energy $\mathcal{F}$ between the states found on cooling
($\mathcal{F}_c$) and on heating ($\mathcal{F}_h$).} \label{fig16}
\end{figure}
In Fig.~\ref{fig16}(a) we show the specific heat capacity vs.
temperature, computed as $c_v = - \frac{T}{V}
\frac{\partial^2\mathcal{F}}{\partial T^2}$. We convert the specific heat capacity from the units of $\mathcal{F}_0/(V \mathrm{K})$ (free energy per volume per Kelvin) to its equivalent SI value which is in our case $6.2 \times 10^{-7} \mathrm{J}\cdot \mathrm{K}^{-1} \cdot \mathrm{cm}^{-3}$ (using  volume of the simulation region $V = 31416$ $\xi_{10}^3 \approx 1.72 \times 10^{-12}$ cm$^3$). The most pronounced feature in the specific heat capacity curve is a jump $\Delta c_{v}$ of approximately 109 nJ $\cdot$ K$^{-1}\cdot$ cm$^{-3}$, where on heating the transition from small clusters to a
single stripe is immediately followed by the rapid transition of the
type-I layer through its own $T_c$ to only proximity induced
superconductivity. After that transition the vortices occupy 
the entire sample evenly, forming an Abrikosov lattice. Therefore we can
associate the onset of the attraction between vortices directly with
the onset of superconductivity in the type-I layer. This implies
that for considered parameters type-I layer must be below its own
critical temperature in order to observe any unusual clustering of
vortices. 

On cooling the heat capacity shows a similar jump, but associated only with the transition from Abrikosov lattice to the stripe phase. The subsequent transition at 0.85 K corresponds to the rearrangement of the vortices from the stripe into 
clusters, it is of first order, and associated with latent heat $L \approx 150$ $\mu$J $\cdot$ cm$^{-3}$. The described features in the heat capacity are
ideal for observing the phase transitions of soft vortex matter by
calorimetry, similarly to what is proposed for detection of flux phases inside the 3D samples and distinction of giant vortex to multivortex transitions in Ref.~\onlinecite{PRLXu}. The experimental
realization of the required high-precision calorimetry is feasible,
and has already been reported in Ref.~\onlinecite{Bourgeois}.

In Fig.~\ref{fig16}(b) we then show the difference in the free energies
$\mathcal{F}_c - \mathcal{F}_h$ between the states found on 
cooling and heating respectively. The sign of this quantity determines which of the
possible states is energetically favorable and details the energy
cost of the metastable states. For example, one can directly see
from Fig.~\ref{fig16}(b) that the transition between the stripe and
clusters in the equilibrium should occur at $T \approx$ 2.8 K.

\section{Conclusions} We presented novel and rich vortex phases
and phase transitions in a type-I/type-II superconducting bilayer,
resembling known phenomena in soft matter physics. The
solution-gel-glass-crystal transitions of vortex matter can be
induced in our system by an external magnetic field, current or
temperature, but can be also engineered by a proper choice of the
constituent materials, thinning the type-I layer to effective
type-II behavior, or changing the spacer material to influence the
coupling strength. The proposed superconducting system is in many
ways peculiar and different from any soft matter system, but
similarities arise from the competing interactions with different
length scales - present in both systems. Our superconducting system is controllable, relatively easy to fabricate and allows for convenient vortex imaging or detection of transitions between phases using neutron scattering or calorimetric measurements.
Moreover, this system opens a further research direction, leaning
upon the early discovery of Giaver that it is possible to make a DC
transformer by using applied current in one superconductor to drag
vortices through another and induce voltage there.\cite{Giaver} The
inability to {\it ad hoc} predict what would happen to soft vortex
matter phases in that case, as well as the links to related studies
of Coulomb drag in semiconductor heterostructures\cite{GaAlAs} and
bilayer graphene\cite{Geim}, make our system a very
interesting testbed for a plethora of new phenomena.

\begin{acknowledgments}
This work was supported by the Flemish Science Foundation (FWO-Vl).
Insightful discussions with Arkady Shanenko and Edith Cristina Euan
Diaz are gratefully acknowledged.
\end{acknowledgments}

\end{document}